\documentclass[12pt]{article}
\usepackage[utf8]{inputenc}
\usepackage[T1]{fontenc}
\usepackage{natbib}
\usepackage{geometry}

 \global\bibsep=0pt

\usepackage[colorlinks=false]{hyperref}

\usepackage{sectsty}

\sectionfont{\normalsize}
\subsectionfont{\small}
\subsubsectionfont{\small}
\paragraphfont{\small}
 
\usepackage{amsmath}
\usepackage{amssymb}
\usepackage{wasysym}
\usepackage{subcaption}
\usepackage{units}
\usepackage{booktabs}
\usepackage{graphicx,color}

\newcommand{\vp}{v_{\mathrm{p}}}
\newcommand{\vs}{v_{\mathrm{s}}}

\newcommand{\vpI}[1]{v_{\mathrm{p}_{#1}}}
\newcommand{\vsI}[1]{v_{\mathrm{s}_{#1}}}
\newcommand{\ms}{\mathrm{m}/\mathrm{s}}
\newcommand{\m}{\mathrm{m}}
\newcommand{\Hz}{\mathrm{Hz}}
\newcommand{\kgm}{\mathrm{kg}/\mathrm{m}^{3}}
\newcommand{\rads}{\mathrm{rad}/\mathrm{s}}

\newcommand{\ii}{\mathrm{i}}

\newcommand{\Hanksz}[1]{H_{0}^{\left( 2 \right)} \!\! \left( #1 \right)}

\newcommand{\Hanksf}[1]{H_{1}^{\left( 2 \right)} \!\! \left( #1 \right)}

\newcommand{\bfd}{\boldsymbol{d}}
\newcommand{\bfe}{\boldsymbol{e}}

\newcommand{\bffhat}{\boldsymbol{\hat{f}}}
\newcommand{\bfm}{\boldsymbol{m}}
\newcommand{\bfs}{\boldsymbol{s}}
\newcommand{\bfx}{\boldsymbol{x}}
\newcommand{\bfu}{\boldsymbol{u}}
\newcommand{\bfuhat}{\boldsymbol{\hat{u}}}
\newcommand{\bfB}{\boldsymbol{B}}
\newcommand{\bfC}{\boldsymbol{C}}
\newcommand{\bfK}{\boldsymbol{K}}
\newcommand{\bfM}{\boldsymbol{M}}
\newcommand{\bfN}{\boldsymbol{N}}
\newcommand{\bfL}{\boldsymbol{L}}
\newcommand{\bfdelta}{\boldsymbol{\delta}}

\title{\normalsize\bf%
\Large{Elastic waveform inversion in the frequency domain\\for an application in mechanized tunneling}
}

\author{\normalsize
Christopher Riedel$^{\mathrm{a},}$\footnote{Corresponding author\newline \quad E-mail address: \texttt{christopher.riedel@rub.de} (Christopher Riedel)} , Khayal Musayev$^{\mathrm{a}}$, Matthias Baitsch$^{\mathrm{b}}$ and Klaus Hackl$^{\mathrm{a}}$
}

\begin{document}
\date{}

\maketitle

\vspace{-1.0cm}

\begin{center}
{\footnotesize 
$^\mathrm{a}$Institute of Mechanics of Materials, Ruhr-Universit\"at Bochum, Bochum, Germany \\
$^\mathrm{b}$Bochum University of Applied Sciences, Bochum, Germany \\
}
\end{center}

\bigskip

\hrule

\bigskip
\noindent
{\small{\bf Abstract}\medskip\\
The excavation process in mechanized tunneling can be improved by reconnaissance of the geology ahead. A nondestructive exploration can be achieved in means of seismic imaging. A full waveform inversion approach, which works in the frequency domain, is investigated for the application in tunneling. The approach tries to minimize the difference of seismic records from field observations and from a discretized ground model by changing the ground properties. The final ground model might be a representation of the geology. The used elastic wave modeling approach is described as well as the application of convolutional perfectly matched layers.
The proposed inversion scheme uses the discrete adjoint gradient method, a multi-scale approach as well as the L-BFGS method.
Numerical parameters are identified as well as a validation of the forward wave modeling approach is performed in advance to the inversion of every example.
Two-dimensional blind tests with two different ground scenarios and with two different source and receiver station configurations are performed and analyzed, where only the seismic records, the source functions and the ambient ground properties are provided.
Finally, an inversion for a three-dimensional tunnel model is performed and analyzed for three different source and receiver station configurations.
}

\bigskip

\hrule

\bigskip

\baselineskip=\normalbaselineskip

\section{Introduction}
\label{sec:introduction}

Different tasks are tackled by the construction of tunnels as the reduction of urban traffic, the improvement of transportation systems and the creation of direct traffic links. 
Mechanized tunneling is an efficient choice for the construction of circular tunnels, where a cylindrical tunnel boring machine (TBM) drills through the ground by using abrasive wear on the rotating front tunnel shield. Many processes are automatized like the installation of the lining and the mucking out of the debris to ensure a fast and safe construction process with a high quality. Despite many advantages, the process has difficulties to adapt to unexpected geological conditions, which may lead to unwanted damages of the TBM as well as dwell times.
A detection of geological changes a few meters in front of the tunnel gives the opportunity to initiate countermeasures, which would lead to a reduction of time, costs and damage.

Seismic exploration has already been used for reconnaissance in tunneling. For this purpose, seismic nondestructive waves are emitted into the ground. These waves are reflected and refracted due to geological changes. The seismic activity is measured at several locations and the resulting seismic records contain information about geological conditions. Already industrialized approaches for revealing these information usually use only onsets from reflected direct waves, as e.g. the `Tunnel Seismic Prediction' method \citep{sattel1996} or the `Soft Sonic Probing' method \citep{kneib2000}. Therefore, they are not using all waveforms of the seismic records for their prediction.

The concept of full waveform inversion uses the full content of measured seismic records. Initially introduced by \citet{tarantola1984,tarantola1986} the method aims to reduce the difference of measured and synthetically calculated seismic records by optimizing the material properties of the computational model. Full waveform inversion has already been applied for many large scale problems \citep{fichtner2011} but not for real-time reconnaissance in mechanized tunneling because the amount of needed forward wave simulations is very high and with the current computational resources too time-consuming. But an application of full waveform inversion seems to become feasible in the future because of the continuous improvement of computational power.  
Extensive investigations of different possible full waveform inversion approaches have to be carried out to ensure a fast and skilled implementation in tunneling projects when its application becomes feasible.

Many full waveform inversion approaches perform the minimization as well as the forward wave modeling in the time domain, which is much more intuitive in its utilization than frequency domain approaches because emerging amplitudes in the seismic records can be easier identified and the used time interval can be accurately chosen. On the other hand, the abundance of information, which is used for full waveform inversion, increases the non-linearity of the inverse problem. Frequency domain approaches allow a separate analysis of low and high frequencies. In combination with an intuitive application of a multi-scale approach (increasing used frequencies gradually during the inversion) a mitigation of the non-linearity can be achieved \citep{ajo-franklin2005}. The application of frequency domain full waveform inversion schemes for different problems have been studied e.g. by \citet{pratt1999a} for the inversion of tomographic seismic data of a physical scale model and by \citet{pratt1999b} for the inversion of seismic data of a crosshole experiment.

The potential of full waveform inversion for exploration in a mechanized tunnel domain have already been investigated for several time domain approaches: \citet{bharadwaj2017} use a combination of an adjoint gradient method and a two-dimensional model, which focuses on horizontally polarized shear waves; \citet{lamert2019} use the adjoint gradient method in combination with a nodel discontinuous Galerkin approach; \citet{nguyen2016} developed a hybrid non-deterministic approach by combining simulated annealing with the unscented Kalman filter, which \citet{trapp2021} have successfully applied on measured data of a small-scale experimental setup.

The potential of frequency domain approaches in the context of mechanized tunneling have only been investigated in combination with the usage of the acoustic wave equation \citep{riedel2021,wang2021,yu2021}, which neglects the dominant influence of shear as well as surface waves.
The proposed approach extends the work which has been presented by \citet{musayev2017} and \citet{riedel2019,riedel2021pamm}. In Comparison to \citet{riedel2021} the elastic wave equation is used instead of the acoustic wave equation for forward and inverse wave modeling, which increases the complexity of the inverse problem.

In \autoref{sec:forwardInverseWaveModeling} the forward and inverse wave modeling approaches, which are used for the full waveform inversion, are explained. A key feature for the fast and physical meaningful computation of the wave fields is the accurate application of convolutional perfectly matched layers, which damp waves at the artificial borders of the considered domain to imitate an infinite domain. The used inversion concepts as well as their application are described.
Two two-dimensional blind tests in a tunnel domain are illustrated in \autoref{sec:blindtest}, where synthetic seismic records are provided without any information of the geological changes. The influence of two different configurations of sources and receivers is investigated for two different ground scenarios. The wave modeling approaches from \autoref{sec:forwardInverseWaveModeling} are validated for the used ambient ground model in advance. 
The inversion of seismic records from a synthetic three-dimensional tunnel environment with an embedded layer change is analyzed in \autoref{sec:layerChange3D}. Again, a validation of the wave modeling for the used ground domain is performed in advance. The influence of three different configurations of sources and receivers on the inversion results is investigated.
Finally, the results are summarized by the conclusion in \autoref{sec:conclusion}.
\section{Forward and inverse wave modeling}
\label{sec:forwardInverseWaveModeling}
The application of an accurate numerical model, which calculates physically meaningful wave fields, is essential for gaining reasonable predictions of the geology of the subsurface from the full waveform inversion process.
\subsection{Frequency domain modeling of elastic waves}
\label{subsec:forwardModeling}
For calculating the complex-valued vectorial displacements of the elastic wave field $\bfu$ in the frequency domain for a specific angular frequency $\omega$ the frequency domain formulation of the elastic wave equation has to be solved:
\begin{equation}
- \omega^{2} \rho \! \left( \bfx \right) \bfu \! \left( \bfx , \omega \right)  - \nabla \cdot \left( \bfC \! \left( \bfx \right) : \nabla \bfu \! \left( \bfx , \omega \right) \right) = f_{\omega} \! \left( \omega \right) \bfdelta \! \left( \bfx - \bfs \right).
\label{eq:elasticWaveEquation}
\end{equation}
The density of the subsurface is indicated by $\rho \left( \bfx \right)$ for the spatial position $\bfx$. For a simplified application, the excitation force has been decomposed in a complex-valued scalar part $f_{\omega} \left( \omega \right)$, which depends on the examined angular frequency $\omega$, as well as in a real-valued vectorial part $\bfdelta \! \left( \bfx - \bfs \right)$, which is a Dirac delta function that only acts at the spatial position of the source $\bfs$ in a defined direction. The elastic properties of the ground are stored in the fourth-order material stiffness tensor $\bfC$ and are described within this study in terms of the real-valued P-wave velocity $\vp$ and S-wave velocity $\vs$.
They are related to the Lam\'{e} parameters $\lambda$ and $\mu$ by the following equations:
\begin{equation}
\vp = \sqrt{\frac{\lambda+2\mu}{\rho}}, \quad \vs = \sqrt{\frac{\mu}{\rho}}.
\label{eq:VpVs}
\end{equation}
The Kolsky-Futtermann model \citep{kolsky1956,futterman1962} can be implemented with low additional effort within the frequency domain approach to take attenuation effects into account. The real-valued elastic properties are replaced by complex-valued counterparts, which are frequency-dependent. The intensity of the attenuation is described by a quality factor for each elastic property. The impact of attenuation effects on the application of full waveform inversion in a tunnel environment has already been investigated for slightly underestimated and overestimated attenuation by \citet{lamert2019} and \citet{riedel2021}. Within this study no attenuation effects are considered.

Certain boundary conditions have to be included for solving \autoref{eq:elasticWaveEquation}. At the tunnel walls as well as at the Earth's surface a Neumann boundary conditions has to be applied:
\begin{equation}
\boldsymbol{\sigma} \cdot \boldsymbol{n} = \boldsymbol{0}.
\label{eq:NeumannBC}
\end{equation}
At these free surfaces the stress $\boldsymbol{\sigma}$ in the normal direction of the ground $\boldsymbol{n}$ vanishes. Seismic waves which reach these surfaces are reflected and induce surface waves which travel along the surface.
The comparatively high amplitudes of the surface waves impede the seismic exploration because amplitudes from body waves, which have been reflected and refracted at material discontinuities, are partially hidden. 

Only a finite domain $\varOmega$ in front and around the tunnel has to be considered for reconnaissance purposes. The computational costs are reduced by an application of artificial boundaries at the edges of the domain of interest. But without high attenuation effects seismic waves would propagate in infinite space. Therefore, unrealistic reflections at the artificial boundaries have to be suppressed. An efficient way is the usage of perfectly matched layers (PMLs), which absorb the escaping waves by a coordinate stretching into the imaginary plane. The original idea has been proposed by \citet{berenger1994} and the approach of \citet{festa2005}, which uses convolutional perfectly matched layers (cPMLs) to deal with numerical instabilities, which occur due to surface waves in shallow domains, is used for this study.

\begin{figure}
\centering
\includegraphics[width=0.45\textwidth]{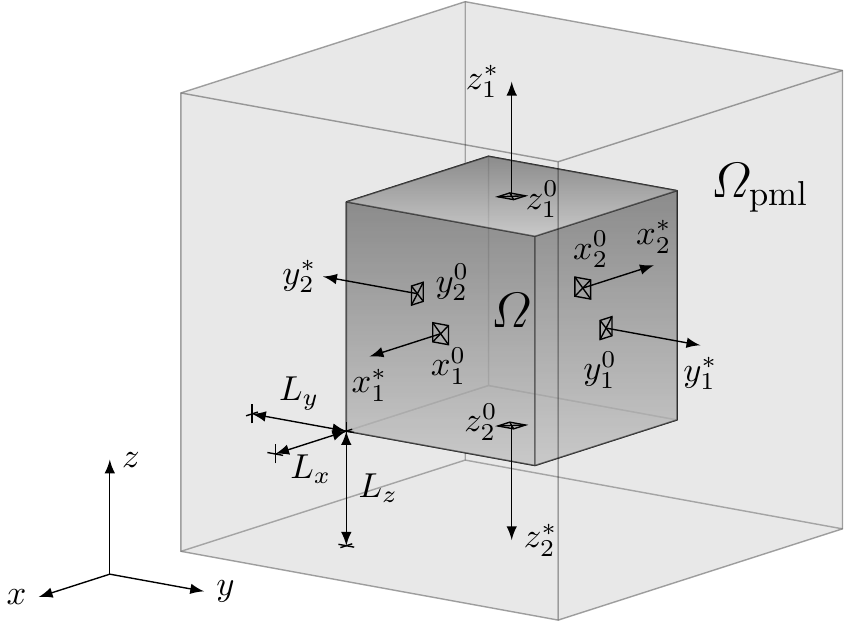}
\caption{A zone of perfectly matched layers $\varOmega_{\mathrm{pml}}$ surrounds a finite domain of interest $\varOmega$ for imitating infinite space.}
\label{fig:PML}
\end{figure}

The application of a convolutional PML in x-direction will be discussed in the same manner as by \citet{riedel2021}. An application in the other spatial directions is straight forward. Within the PML domain $\varOmega_{\mathrm{pml}}$ the real-valued coordinate $x$ is replaced by the complex-valued coordinate $\tilde{x}$: 
\begin{equation}
	\tilde{x} = x^{0}_{e} + \int_{x^{0}_{e}}^{x} \varepsilon_{x} \! \left( x_{e}^{*} \! \left( x^{\prime} \right) \right) \mathrm{d} x^{\prime},
	\label{eq:PML_complexCoordinate}
\end{equation}
where the stretching function $\varepsilon_{x}$ describes the coordinate stretching into the imaginary plane:
\begin{equation}
	\varepsilon_{x} \! \left( x^{*}_{e} \right) = 1 + \frac{\gamma_{x} \! \left( x^{*}_{e} \right)}{\omega_{\mathrm{c}} + \mathrm{i} \omega}.
	\label{eq:PML_epsilon}
\end{equation}
A local coordinate $x^{*}_{e} = \left| x - x^{0}_{e} \right|$ is used to describe the current position within the layer with respect to the different inner edges $e$ (see \autoref{fig:PML}).
The damping function
\begin{equation}
	\gamma_{x} \! \left( x^{*}_{e} \right) = c_{\mathrm{pml}} \left( 1 - \cos \left( \frac{\pi}{2} \frac{x}{L_{x}} \right) \right),
\end{equation}
starts from zero and strictly increases to the parameter $c_{\mathrm{pml}}$ at the end of the layer to guarantee a smoothed absorption behavior.
The performance of the PML depends on the chosen width $L_{x}$, the used discretization and especially on an accurate evaluation of the parameter $c_{\mathrm{pml}}$.
In comparison to standard PML the parameter $\omega_{\mathrm{c}}$ has been added to the denominator in \autoref{eq:PML_epsilon}. As stated by \citet{festa2005_SEM} the additional term acts as a Butterworth filter. Within this study $\omega_{\mathrm{c}} = 0.99 \cdot \omega$ is used.

The coordinate stretching affects \autoref{eq:elasticWaveEquation} in terms of the partial derivatives. Therefore, the partial derivative with respect to the complex coordinate $\tilde{x}$ is evaluated using the chain rule as well as the inverse function theorem:
\begin{equation}
\frac{\partial}{\partial \tilde{x}} = \frac{\partial x}{\partial \tilde{x}} \frac{\partial}{\partial x} = \varepsilon_{x}^{-1} \frac{\partial}{\partial x}.
\label{eq:partDerivComplexX}
\end{equation}
Outside the PMLs the stretching functions $\varepsilon_{x}$, $\varepsilon_{y}$ and $\varepsilon_{z}$ are equal one. Therefore, an adoption of the elastic wave equation (\autoref{eq:elasticWaveEquation}) is performable without changing its application in the investigated domain $\varOmega$, which is shown here for the three-dimensional case (cf. \citet{zheng2002}):
\begin{equation}
- \omega^{2} \varepsilon_{x} \varepsilon_{y} \varepsilon_{z} \rho \! \left( \bfx \right) \bfu \! \left( \bfx , \omega \right)  - \nabla \cdot \left( \tilde{\bfC} \! \left( \bfx \right) : \nabla \bfu \! \left( \bfx , \omega \right) \right) = f_{\omega} \! \left( \omega \right) \bfdelta \! \left( \bfx - \bfs \right).
\label{eq:PMLelasticWaveEquation}
\end{equation}
This smart representation is achieved by replacing the material stiffness tensor $\bfC$ by the complex counterpart $\tilde{\bfC}$, which is usually not isotropic anymore within the PMLs. By using Einstein sum notation, $\tilde{\bfC}$ is defined by:
\begin{equation}
\begin{split}
\tilde{C}_{ijkl} &= \frac{\varepsilon_{x} \varepsilon_{y} \varepsilon_{z}}{\varepsilon_{i} \varepsilon_{k}} C_{ijkl}\\
&= \frac{\varepsilon_{x} \varepsilon_{y} \varepsilon_{z}}{\varepsilon_{i} \varepsilon_{k}} \rho \left( \left( \vp^{2} - 2 \vs^{2} \right) \delta_{ij} \delta_{kl} + \vs^{2} \left( \delta_{il} \delta_{jk} + \delta_{ik} \delta_{jl} \right) \right),
\end{split}
\label{eq:PMLmaterialStiffnessTensor}
\end{equation}
where $\delta_{ij}$ is the Kronecker delta, which yields one if both indices are equal and zero if not.

A finite element approach, which utilizes higher-order hierarchical shape functions \citep{szabo2004}, is applied to solve \autoref{eq:PMLelasticWaveEquation}. The advantage of using hierarchical shape functions instead to higher-order Lagrangian shape functions is that additional higher-order shape functions can be simply added for higher frequencies, which cause more complex wave fields, without changing the finite element mesh.

For the discretized representation of the P-wave velocity $\vpI{\mathrm{h}} \! \left( \bfm , \bfx \right)$ and the S-wave velocity $\vsI{\mathrm{h}} \! \left( \bfm , \bfx \right)$ of the ground a linear combination of basis functions $\varphi_{k}$ and scalar-valued parameters of the material ground properties are used:
\begin{align}
\vpI{\mathrm{h}} \! \left( \bfm , \bfx \right) &= {\displaystyle \sum^{N_{\mathrm{m}}}_{k=1}} m_{k} \, \varphi_{k} \! \left( \bfx \right), \label{eq:matParameterVp}\\
\vsI{\mathrm{h}} \! \left( \bfm , \bfx \right) &= {\displaystyle \sum^{N_{\mathrm{m}}}_{k=1}} m_{N_{\mathrm{m}}+k} \, \varphi_{k} \! \left( \bfx \right), \label{eq:matParameterVs}\\
\bfm &= \left( \vpI{1}, \vpI{2}, \cdots , \vpI{N_{\mathrm{m}}}, \vsI{1}, \vsI{2}, \cdots , \vsI{N_{\mathrm{m}}}  \right)^{\mathrm{T}}.\label{eq:matParameter}
\end{align}
Within this study the functions $\varphi_{k}$ are piecewise linear polynomials and therefore, the number of used nodes is equal $N_{\mathrm{m}}$. The coefficient vector $\bfm$ stores all parameters of the elastic material properties of the ground.
\autoref{eq:matParameterVp}, \autoref{eq:matParameterVs} and \autoref{eq:matParameter} could be extended to take the spatial variation of the density $\rho \! \left( \bfx \right)$ into account, which is not necessary for this study because the density is considered to be constant.

In terms of finite element analysis \citep{szabo2011} the complex-valued displacement coefficients as well as external force coefficients are condensed in the vectors $\bfuhat$ and $\bffhat$. The Stiffness matrix $\bfK$ as well as the mass matrix $\bfM$ can be introduced by using the shape function matrix $\bfN$, the differential operator matrix $\bfB$ as well as the complex material stiffness matrix $\tilde{\bfC}_{\mathrm{Voigt}}$ ($\tilde{\bfC}$ in Voigt notation): 
\begin{equation}
\begin{split}
\bfK &= - \int_{\varOmega} \bfB^{\mathrm{T}} \!\! \left( \bfx \right) \tilde{\bfC}_{\mathrm{Voigt}} \! \left( \bfx \right) \bfB \! \left( \bfx \right) \mathrm{d}\varOmega,\\
\bfM &= \int_{\varOmega} \varepsilon_{x} \! \left( \bfx \right) \varepsilon_{y} \! \left( \bfx \right) \varepsilon_{z} \! \left( \bfx \right) \rho \bfN^{\mathrm{T}} \!\! \left( \bfx \right) \bfN \! \left( \bfx \right) \mathrm{d}\varOmega,\\
\bffhat &= f_{\omega} \! \left( \omega \right) \bfN^{\mathrm{T}} \!\! \left( \bfs \right) \bfdelta \! \left( \bfx - \bfs \right).
\end{split}
\label{eq:MatricesFEM}
\end{equation}
The Neumann boundary condition (\autoref{eq:NeumannBC}) is implicitly fulfilled.
The discretized elastic wave equation in the frequency domain turns out to be a linear system of equations:
\begin{equation}
\left( -\omega^{2} \bfM + \bfK \right) \bfuhat = \bfL \bfuhat = \bffhat.
\label{eq:EquilibriumFEM}
\end{equation}
The impedance matrix $\bfL$ conflates the stiffness matrix $\bfK$, the mass matrix $\bfM$ as well as the angular frequency $\omega$.
In comparison to time domain modeling, additional source excitations can be modeled in the frequency domain with low additional computational effort by reusing the factorization of the impedance matrix $\bfL$ of a direct solver.  

The evaluation of the parameter $c_{\mathrm{pml}}$ and the validation of the forward wave modeling approach is performed in \autoref{sec:blindtest} and in \autoref{sec:layerChange3D} for the analyzed examples.

\subsection{Full waveform inversion}
\label{subsec:inverseModeling}

Seismic nondestructive waves are emitted into the ground by a chosen number of sources $N_{\mathrm{s}}$ at different locations to illuminate the ground around and in front of the tunnel. The propagating waves travel through the ground and are reflected as well as converted into surface waves at the free surfaces. Reflections as well as refractions of the elastic waves occur at the interfaces of geological changes. A number of installed receivers $N_{\mathrm{r}}$, which have been placed at convenient locations, record the seismic activity. The measured seismic records contain information of the ground properties due to the influence of seismic reflections and refractions at the interfaces of geological changes. The locations of sources and receivers affect which and how much information of the different ground formations are gathered within the seismic records. Therefore, a wide spreading of sources and receivers would improve the exploration. In practice their numbers and locations are constrained by economical objectives and by the access to the locations.

The full waveform inversion procedure tries to predict the geological conditions of the subsurface by reducing the difference of the measured seismic records and seismic records of a numerical model of the ground by adapting the ground properties of this model. The ground properties of the final synthetic ground model shall be a rough representation of the real ground conditions.

A misfit functional $\chi$ has to be defined as objective functional for this minimization problem. The application of a multi-scale approach is very common in the framework of frequency domain full waveform inversion for reducing the non-linearity of the problem (e.g. \citet{pratt1999a}).
Therefore, the misfit $\chi_{i}$ is minimized successively for $N_{\mathrm{fg}}$ frequency groups $F_{i}$, which contain $N_{\mathrm{f}}$ different angular frequencies $\omega_{if}$:
\begin{equation}
\begin{split}
F_{i}  = \left\{ \omega_{i1}, \omega_{i2}, \cdots, \omega_{iN_{\mathrm{f}}} \right\}\\
\mathrm{with} \quad i = 1, \cdots, N_{\mathrm{fg}},
\end{split}
\label{eq:frequencyGroup}
\end{equation}
where the highest frequency of a group is always higher than the one of the previous frequency group. 
The resulting coefficients of the ground properties $\bfm$ after the inversion of a frequency group are used as the initial ground model of the following frequency group.
Since the complexity of the misfit functional is proportional to the dominant length scale of the ground model, it is very likely that the current ground model is for low frequencies in the basin of attraction of the global minimum of $\chi$. Using this minimum for higher frequencies increases the probability of being again in the basin of attraction of the global minimum for a more complex misfit functional \citep{fichtner2011}.
This continuous transition from long to short period data leads to an emergence of rough structures at the beginning of the inversion process which get sharper with increasing frequency. 
\citet{brossier2010} have already analyzed the performance of different data residual norms in the context of elastic frequency domain full waveform inversion for synthetic offshore and onshore examples. As only synthetic examples without noise are investigated in this study, the least squares norm seems to be the best option:
\begin{equation}
\begin{split}
\chi_{i} \! \left( \bfm \right) = {\displaystyle \sum\limits_{f=1}^{N_{\mathrm{f}}} \sum\limits_{s=1}^{N_{\mathrm{s}}} \sum\limits_{r=1}^{N_{\mathrm{r}}} \sum\limits_{d=1}^{N_{\mathrm{d}}}} \left( \Delta u_{fsrd} \! \left( \boldsymbol{m} \right) \right) \left( \Delta u_{fsrd} \! \left( \boldsymbol{m} \right) \right)^{*}\\
\mathrm{with} \quad \Delta u_{fsrd} \! \left( \boldsymbol{m} \right) = u_{fsrd} \! \left( \boldsymbol{m} \right) - u_{fsrd}^{0}.
\end{split}
\label{eq:misfit}
\end{equation}
The operator $\left(  \bullet \right)^{*}$ denotes the complex conjugate. The complex displacement $u$ for a frequency $f$ for a source excitation $s$ at a receiver station $r$ in the direction $d$ depends obviously on the current ground model $\bfm$ in contrast to the complex displacement $u^{0}$ of the seismic records from field observations. 

For updating the ground properties in a way to minimize \autoref{eq:misfit} the discrete adjoint gradient method is used \citep{fichtner2011}. The partial derivative of the misfit for a single frequency of a frequency group with respect to the $k^{\mathrm{th}}$ entry of the coefficient vector $\bfm$ can by evaluated by using the chain rule as well as \autoref{eq:EquilibriumFEM} as an additional constraint:
\begin{equation}
\frac{\partial \chi_{if} \! \left( \bfm, \omega_{if} \right)}{\partial m_{k}} = \bfuhat \cdot \frac{\partial \bfL \! \left( \bfm, \omega_{if} \right)}{\partial m_{k}} \bfuhat^{\dagger}.
\label{eq:gradientAdjoint}
\end{equation}
The coefficient vector of the complex adjoint wavefield $\bfuhat^{\dagger}$ is calculated with very low computational effort by reusing the factorization of a direct solver of the impedance matrix $\bfL$ from the former forward simulation:
\begin{equation}
\bfuhat^{\dagger} = - \left( \bfL\! \left( \bfm, \omega_{if} \right) \right)^{-1} \nabla_{u} \chi_{if},
\label{eq:adjointWaveField}
\end{equation}
where $-\nabla_{u} \chi_{if} = -\partial \chi_{if} /\partial \bfuhat$ acts as the adjoint source that excites at the receiver stations. The gradient vector
\begin{equation}
\begin{split}
\nabla_{m} \chi_{i} \! \left( \bfm \right) = {\displaystyle \sum\limits_{f=1}^{N_{\mathrm{f}}}} \, \frac{\partial \chi_{if} \! \left( \bfm \right)}{\partial m_{k}} \bfe_{k}\\
\mathrm{with} \quad \bfe_{k} = \left( 1,\cdots,1\right)^{\mathrm{T}},
\end{split}
\label{eq:gradientVector}
\end{equation}
stores the single gradient entries, which are accumulated for all frequencies of the current frequency group.
Every entry of the gradient vector is normalized on the associated volume to use pre-integrated expressions (cf. \citep{lamert2019}).

The coefficients of the ground properties are updated for every iteration $j$ of a frequency group:
\begin{equation}
\bfm_{j+1} = \bfm_{j} + \alpha_{j} \bfd_{j}.
\label{eq:updateGroundModel}
\end{equation}
The vector $\bfd_{j}$ denotes the used search direction. The negative gradient is used as search direction for the first iteration of a frequency group $d_{0} = -\nabla_{m} \chi_{i} \! \left( \bfm_{0} \right)$. For the following iterations the L-BFGS method \citep{nocedal2006} is applied to calculate the search direction $\bfd_{j}$, which ensures faster convergence by taking information of the previous iterations into account.
The step length $\alpha_{j}$ has to be determined in a way that it minimizes $\chi_{i} \! \left( \bfm_{j} + \alpha_{j} \bfd_{j} \right)$ within a suitable range around the current ground properties $\bfm_{j}$. In this study a quadratic approximation of the misfit $\chi$ using three points is applied \citep{sun2006}. A suitable step length $\alpha_{j}$ is evaluated by iteratively optimizing the three step lengths of the approximation. 

Additional to the multi-scale approach other regularization strategies are applied to ensure convergence to a physical meaningful prediction of the ground properties. The minimization of a frequency group is terminated after a specified number of iteration to avoid getting stuck in a local minimum even if the actual convergence criteria is not fulfilled yet \citep{fichtner2011}.

Disproportionately high changes of the gradient arise at the locations of the source and receiver stations because of the singularity of the wave field and the adjoint wave field at the position of the applied forces \citep{pratt1990a}, which inhibits the illumination of the domain in front of the tunnel.
Furthermore, the ground at the locations of the source and receiver stations as well as at the tunnel and at the Earth's surface is usually accessible and therefore, its properties are known. To suppress unrealistic high amplitudes of the properties at these positions a preconditioning of the gradient is used by setting the gradient around the free surfaces as well as at the source and receiver stations equal zero. Between these areas and the other domain a small transition zone is applied, which scales the gradient gradually up from zero to is actual value (cf. \citet{lambrecht2015}).
The choice of the initial model of the ground properties is very crucial for the success of the inversion process. Using a model, which is not close enough to the real subsurface, leads to non meaningful results because the process is likely to get stuck in a local minimum. Therefore, information from preliminary investigations of the ground are very important for characterizing the ambient ground properties.
\section{Elastic two-dimensional blind tests}
\label{sec:blindtest}
Since seismic observations in tunneling today are rarely performed with the application of full waveform inversion in mind, most seismic records are either not suitable or not available. Therefore, the described approach is applied to different synthetic mechanized tunneling examples.
The synthetic seismic records, which are used as field observations, are generated by using SPECFEM2D and SPECFEM3D Cartesian \citep{komatitsch2002a,komatitsch2002b,tromp2008,xie2014}, which are published under the CeCILL v2 license as well as under the GPL 2 license and apply the spectral-element method. On the one hand, a different numerical error is produced by this method in contrast to the proposed approach which makes an inversion more challenging. On the other hand, the seismic records are provided in the time domain like real field observations. A discrete Fourier transformation \citep{williams1999} is performed to obtain the frequency domain reference data for the inversion frequencies.

\citet{riedel2021} have already used two-dimensional blind tests (with an application of the acoustic wave equation) to imitate more realistic tunnel exploration circumstances where no information except of the seismic records, the source function and the ambient ground properties are known.
For the same purpose, two-dimensional blind tests with an application of the elastic wave equation have been performed within the framework of the Collaborative Research Center SFB 837 ``Interaction modeling in mechanized tunneling'' using an adjoint time domain approach as well as the proposed frequency domain approach.
The seismic records of four scenarios in a shallow tunnel domain, where different geological changes in front of the tunnel exist, were given for four different source and receiver station configurations.
The results of the time domain approach are already published by \citet{lamert2020}.
Within this study the results of two ground scenarios for two different placements of source and receiver stations of the frequency domain approach are evaluated and interpreted.

\subsection{Tunnel environment}
\label{subsec:blindtestDomain}

For the two-dimensional tunnel domain a tunnel with a diameter of $6\,\m$ whose top surface is located $15\,\m$ below the Earth's surface is considered. Only the last $20\,\m$ of the tunnel are represented. Below the tunnel $15\,\m$ as well as $80\,\m$ in front of the tunnel are investigated.
At all borders convolutional PMLs with a width of $3\,\m$ are applied, except for the tunnel faces and the Earth's surface. 
The influence of the tunnel boring machine, the tunnel lining and the annular gap mortar are neglected in this study.
The ambient ground properties for this case study are $\vp = 4000\,\ms$, $\vs = 2400\,\ms$ and $\rho = 2500\,\kgm$.
\begin{figure}
\centering
\includegraphics[width=0.5\textwidth]{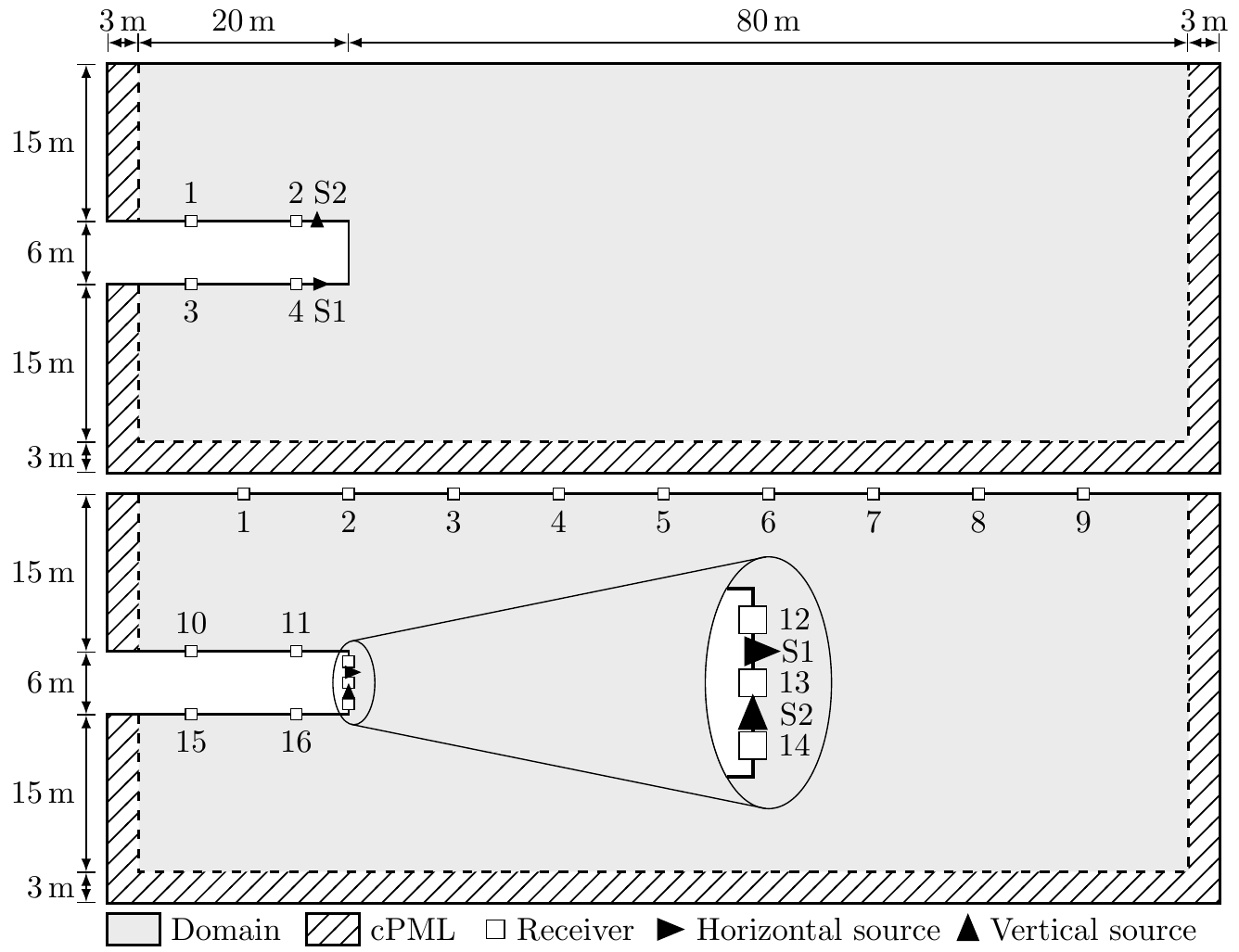}
\caption{Source and receiver stations setup 1 (top) and setup 2 (bottom) that are used for the blind tests.}
\label{fig:SR_Setup}
\end{figure}

The first source and receiver station setup consists of two sources -- one excites in horizontal direction at the tunnel bottom and the other one excites in vertical direction at the ceiling of the tunnel -- and four receiver stations, which are located at the bottom and at the ceiling of the tunnel behind the sources.
Such a setup is likely for today's reconnaissance approaches in tunneling. Tunnel surface waves are initiated which transform to body shear waves at the front of the tunnel and propagate into the investigated domain. Returning reflected waves are transformed at the tunnel to tunnel surface waves whose amplitudes are much higher than body waves and therefore have a better signal to noise ratio \citep{jetschny2010}. This advantage is especially used by the `Tunnel Seismic Prediction' method \citep{sattel1996} and by the `Integrated Seismic Imaging System' \citep{bohlen2003}. The `Tunnel Seismic Prediction' method has been applied in several today's tunneling projects (cf. \citet{lu2015}).

For other seismic exploration techniques in today's tunneling sources and receivers are embedded into the front tunnel shield like for the `Soft Sonic Probing' method \citep{kneib2000}. The second used source and receiver station configuration uses additional to two sources (horizontal and vertical excitation) and three receivers at the front tunnel shield nine receiver stations at the Earth's surface and two receiver stations at the tunnel's bottom as well as at the tunnel's ceiling.
Not only reflected but additionally refracted waves are captured by the receivers due to the dispersion of the receiver stations over the domain. Therefore, an improved prediction of the ground formations is expected.

All receiver stations are measuring the displacements in horizontally as well as in vertical direction. For all sources a Ricker wavelet with a peak frequency of $500\,\Hz$ and no time delay is utilized.
The domain's dimensions as well as the positions and numbering of the source and receiver stations are illustrated in \autoref{fig:SR_Setup}. For the discretization of the tunnel domain $3996$ quadrilateral elements with an area of $1\,\m \times 1\,\m$ are used. 

\subsection{Evaluation of the PML parameter}
\label{subsec:blindtestPML}
As already mentioned in \autoref{subsec:forwardModeling} an accurate evaluation of the PML parameter $c_{\mathrm{pml}}$ for the used models is necessary to ensure a suitable performance of the PMLs. The Green's function $g_{x}$ indicates the impulse response of the system in $x$-direction. The analytical Green's function in $x$-direction $g_{x}^{\mathrm{A}}$ for an angular frequency $\omega$ and for a source excitation in vertical direction within an unbounded domain is given by \citep{min2000}:
\begin{equation}
\begin{split}
g_{x}^{\mathrm{A}} \! \left( r , \theta, \omega \right) = ~~
& \dfrac{\ii}{4 \rho \vp^{2}} \cos \left( \theta \right) \sin \left( \theta \right) \Hanksz{ r \dfrac{\omega}{\vp} } \\ 
-& \dfrac{\ii}{4 \rho \vs^{2}} \cos \left( \theta \right) \sin \left( \theta \right) \Hanksz{ r \dfrac{\omega}{\vs} } \\ 
-& \dfrac{\ii}{2 \rho \vp} \dfrac{\cos \left( \theta \right) \sin \left( \theta \right)}{r \omega} \Hanksf{ r \dfrac{\omega}{\vp} } \\ 
+& \dfrac{\ii}{2 \rho \vs} \dfrac{\cos \left( \theta \right) \sin \left( \theta \right)}{r \omega} \Hanksf{ r \dfrac{\omega}{\vs} } \\
\mathrm{with} \quad & r = \sqrt{\left( x-x_{\mathrm{s}} \right)^{2} + \left( y-y_{\mathrm{s}} \right)^{2}}\\
\mathrm{and} \quad &\theta = \arctan \left( \dfrac{x-x_{\mathrm{s}}}{-(y-y_{\mathrm{s}})} \right).
\end{split}
\label{eq:analyticX}
\end{equation}
The Hankel function of the second kind of zero order $H_{0}^{\left( 2 \right)}$ as well as the Hankel function of the second kind of the first order $H_{1}^{\left( 2 \right)}$ are complex-valued and the position of the used source is indicated by $\left( x_{\mathrm{s}} , y_{\mathrm{s}} \right)$.

\begin{figure}
\centering
\includegraphics[width=0.5\textwidth]{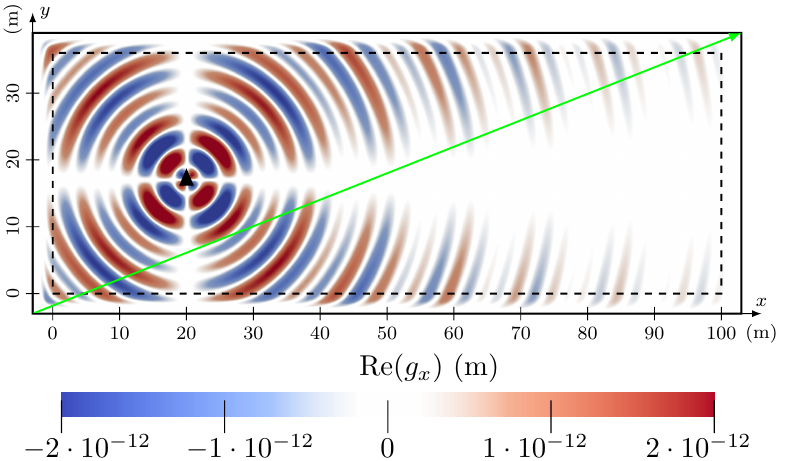}
\caption{Real part of numerical Green's function $g_{x}$ in $x$-direction within a homogeneous domain using PMLs at all borders to mimic infinite space.}
\label{fig:unboundedDomain}
\end{figure}

\begin{figure}
\centering
\includegraphics[width=0.48\textwidth]{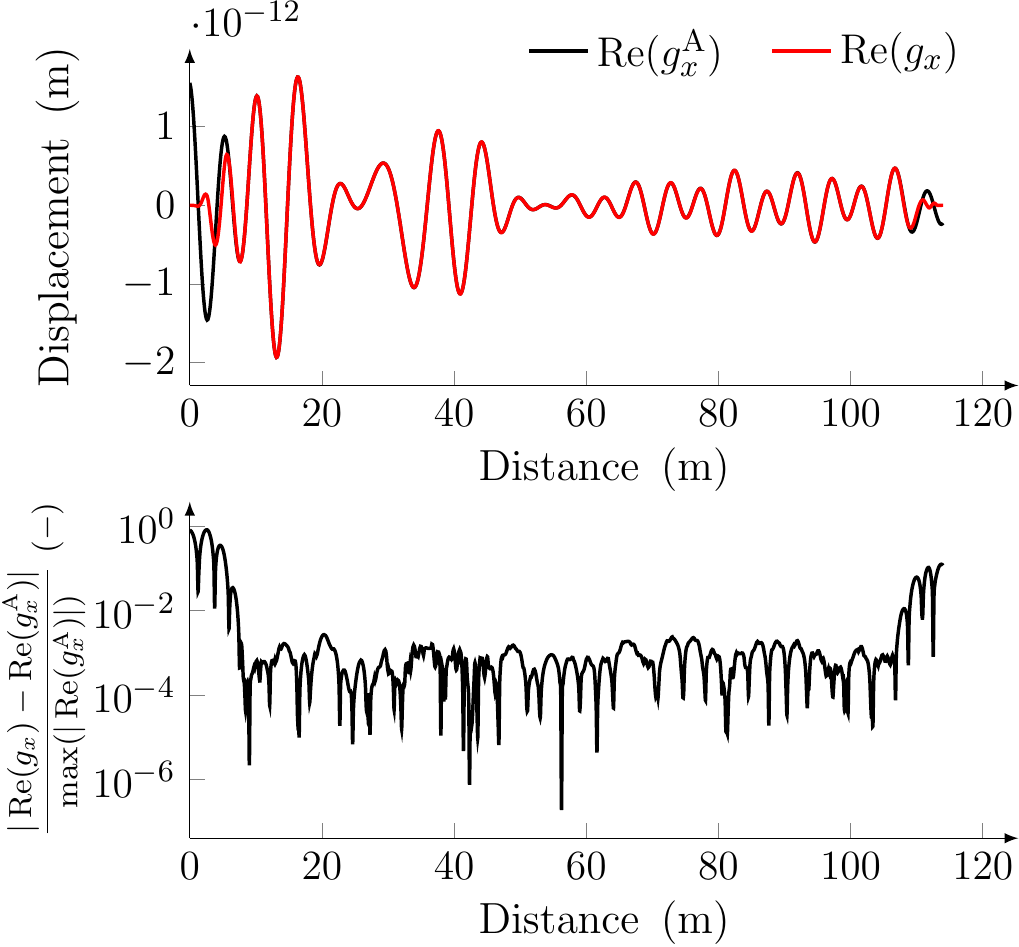}
\caption{Comparison of the numerical and the analytical real part of the Green's function as well as their relative difference along the line indicated in \autoref{fig:unboundedDomain}.}
\label{fig:blindtestPlotOverLine}
\end{figure}

An identification of the parameter $c_{\mathrm{pml}}$ as well as an estimation how many higher order shape functions have to be added for the considered angular frequency $\omega$ to ensure an accurate representation of the wave field can be achieved by comparing the analytical solution of \autoref{eq:analyticX} with the numerical solution of an unbounded domain. This is done for a frequency of $500\,\Hz$ and the real part of the numerical solution of the Green's function in $x$-direction $g_{x}$ is illustrated in \autoref{fig:unboundedDomain}.
The used unbounded domain is equivalent to the used tunnel domain except for the missing tunnel and the Earth's surface, which has been exchanged by another PML. The same discretization as well as the same ground properties are used to ensure that the findings are transferable to the analyzed tunnel domain.
The position of the used vertical source station is equivalent to the second source $\mathrm{S}2$ of the second setup in \autoref{fig:SR_Setup}.
The wanted decrease of the waves inside the PMLs is observed by using $c_{\mathrm{pml}} = 25\,000$ in combination with a polynomial degree of $p = 3$ of the higher-order hierarchical shape functions.
In \autoref{fig:blindtestPlotOverLine} the real part of the numerical and analytical solution of the Green's function in $x$-direction as well as their relative difference are compared along the diagonal line which is indicated in \autoref{fig:unboundedDomain}.
The difference of the numerical and analytical Green's function is visually not identifiable in \autoref{fig:blindtestPlotOverLine} (top) except for the PML regions where the numerical Green's function decays to zero. The logarithmic representation of the relative difference (\autoref{fig:blindtestPlotOverLine} bottom) shows that the numerical error is very small outside the PMLs and that therefore an accurate computation of the wave field is achieved.
The Green's function in $y$-direction as well as the imaginary parts have been investigated, too, but are not illustrated in this study.

\subsection{Validation of elastic wave propagation in the introduced tunnel environment}
\label{subsec:validationTunnelDomain}
\begin{figure}
\centering
\includegraphics[width=0.5\textwidth]{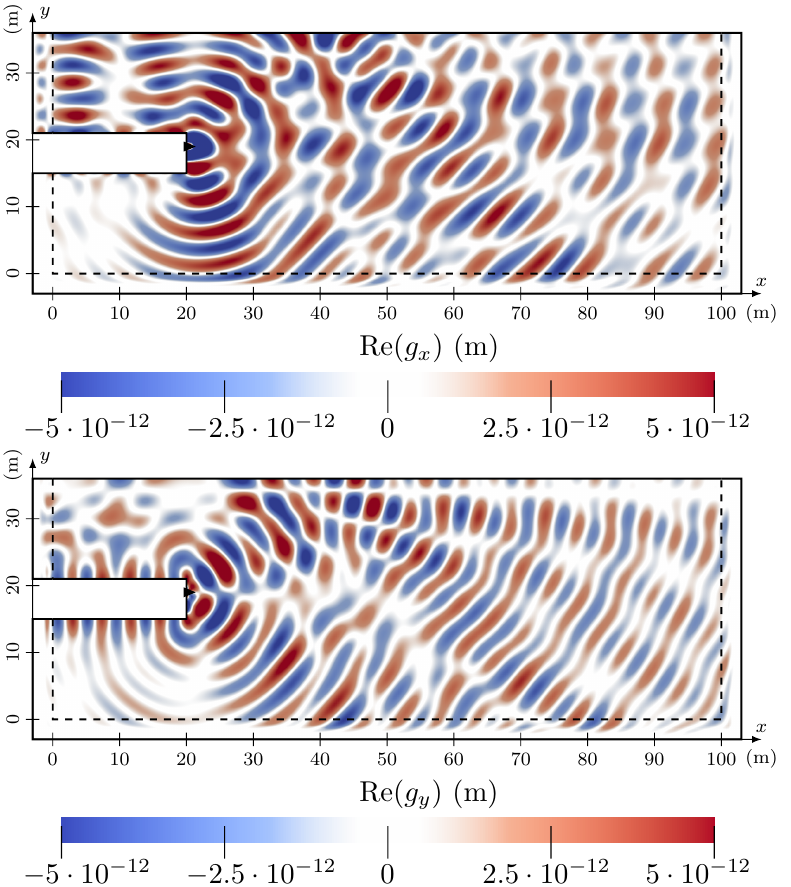}
\caption{Real part of the green's functions $g_{x}$ (top) and $g_{y}$ (bottom) for the used tunnel domain.}
\label{fig:blindtestValidationForwardSimulation}
\end{figure}

The frequency domain wave field of the proposed tunnel domain for an excitation of source station $\mathrm{S}1$ of the second setup for a frequency of $500\,\Hz$ is computed by using the identified parameter $c_{\mathrm{pml}}$. The real part of the displacements are illustrated in \autoref{fig:blindtestValidationForwardSimulation}.
Whether the effects of the reflecting surface are modeled in a physical meaningful way or not is not rateable by this wave field.
But the occurring high amplitudes within the PML between the tunnel and the Earth's surface, which are not decay like the waves in the other PMLs, are conspicuous.

\begin{figure}
\centering
\includegraphics[width=0.495\textwidth]{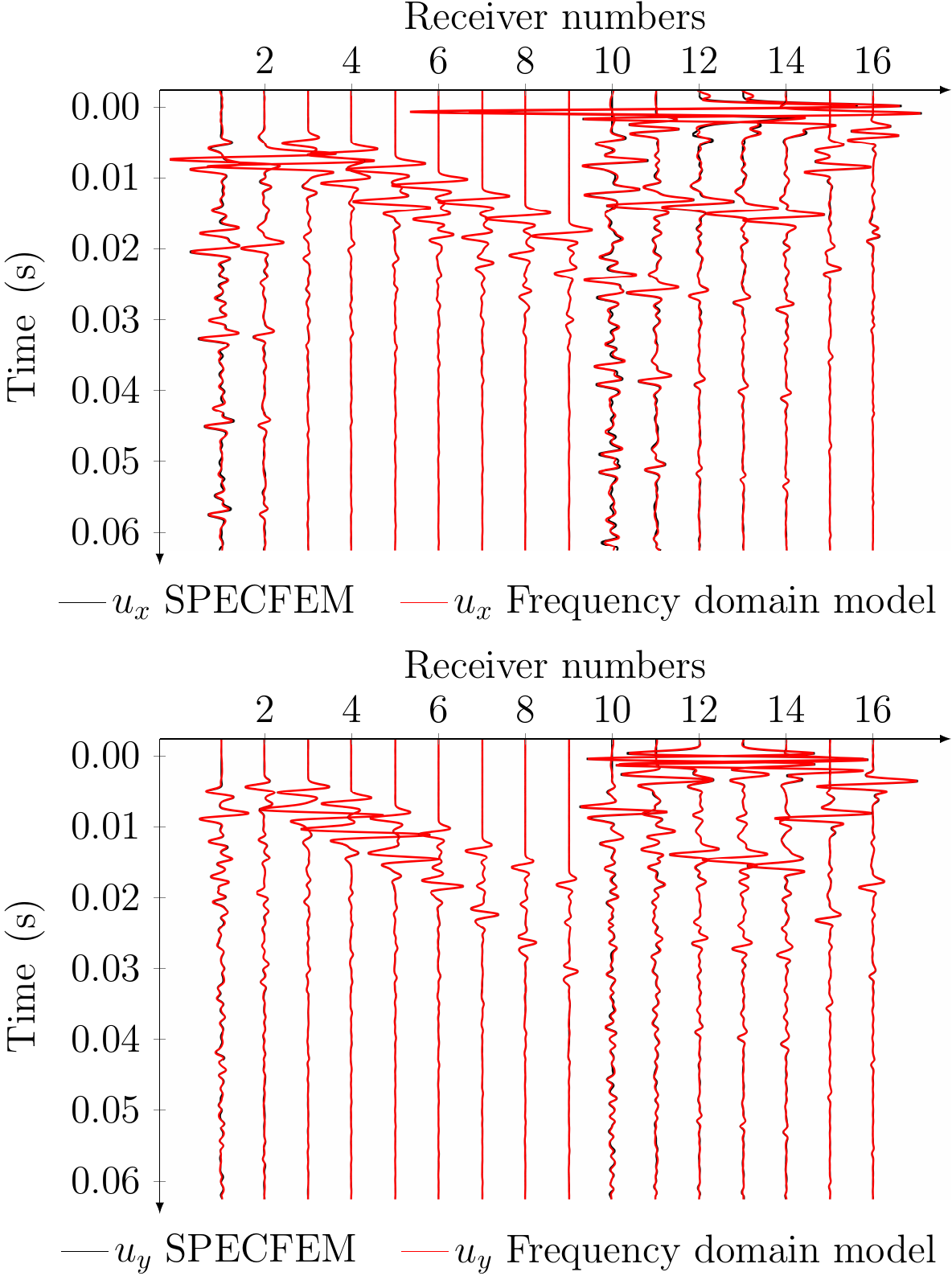}
\caption{Comparison of the seismic records of the proposed frequency domain model and SPECFEM2D for an excitation of source $\mathrm{S}1$ of the second source and receiver station configuration of \autoref{fig:SR_Setup}.}
\label{fig:ValidationBlindtestSetup2H}
\end{figure}

\begin{figure}
\centering
\includegraphics[width=0.5\textwidth]{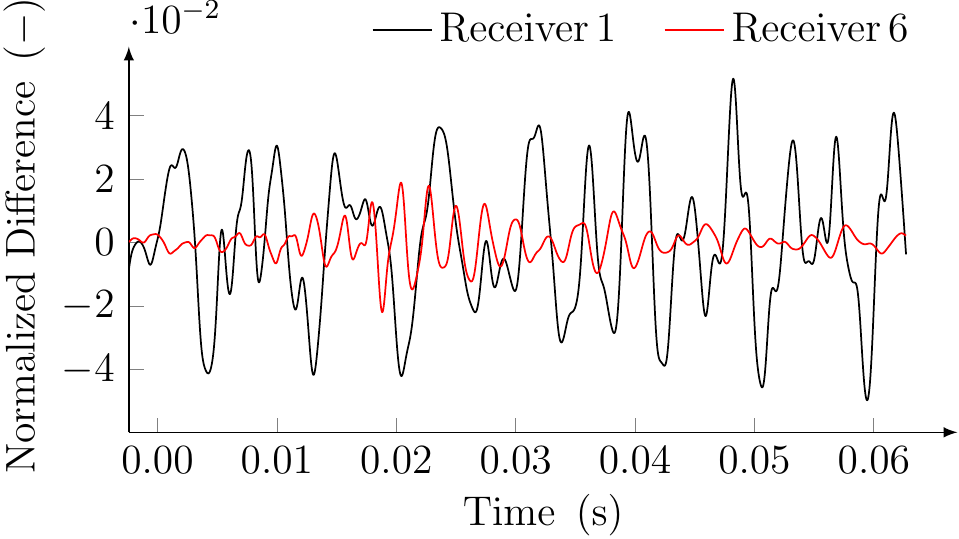}
\caption{Difference of vertical seismic records of the frequency domain model and SPECFEM2D of receiver station 1 and receiver station 6, which are normalized by the maximum amplitude of the corresponding seismic record from SPECFEM2D.}
\label{fig:ValidationBlindtestSetup2HDiffR1R6}
\end{figure}

To check whether physically wave fields are approximated in a tunnel environment by the proposed approach or not, the time domain representation of the seismic records at the receiver stations of the second setup (\autoref{fig:SR_Setup} bottom) for an excitation of source station $\mathrm{S}1$ is compared in \autoref{fig:ValidationBlindtestSetup2H} with seismic records from SPECFEM2D for homogeneous ground conditions. Therefore, the Green's functions at the receiver stations are computed for a frequency range from $100\,\rads$ to $9000\,\rads$ in steps of $10\,\rads$. The degree of the higher-order shape functions is increased with increasing frequencies. These frequency domain seismic records are convolved with a Ricker wavelet with a peak frequency of $500\,\Hz$ and transformed into the time domain using discrete inverse Fourier transformation. 
The seismic records fit overall quiet good together. Small numerical errors occur at the receiver stations next to the used source. Furthermore, the seismic records between the Earth's surface and the tunnel show small differences, too. The normalized difference of the seismic records in vertical direction at receiver station $1$ and $6$ are compared in \autoref{fig:ValidationBlindtestSetup2HDiffR1R6} to illustrate this conspicuousness.   
The differences at receiver station $1$ are much higher.
An explanation for that would be that since an infinite time is modeled in the frequency domain an endless reflection of the seismic waves between the Earth's surface and the tunnel's ceiling is simulated.
This effect is enhanced, on the one hand, by not considering attenuation and, on the other hand, by the application of a two-dimensional model where symmetry is assumed in the third dimension. This effect would not occur in a three-dimensional tunnel domain because the cylindrical tunnel shape would cause that the reflected waves are spread in all directions. This conclusion would explain the not decaying waves within the PML between Earth's surface and tunnel ceiling in \autoref{fig:blindtestValidationForwardSimulation}.
Nevertheless, the proposed wave modeling approach allows a computation of accurate as well as physical meaningful wave fields in a mechanized tunneling domain.

\subsection{Full waveform inversion results}
\label{subsec:resultsBlindtests}
For the inversion of the given seismic records a set $G$ of 28 frequency groups is used. Eight groups contain only a single frequency and 20 groups contain two frequencies. The higher frequency within a group is continuously increasing from group to group, while the other tries to preserve comparatively low frequency content:
\begin{equation}
\begin{split}
G = \left\{ \right. \! & \left\{ 300 \right\}, \left\{ 400 \right\}, \cdots , \left\{ 1000 \right\}, \\
 \! & \left\{ 600,1200 \right\}, \left\{ 700,1400 \right\}, \cdots ,\\
 \! & \left\{ 2100,4200 \right\}, \left\{ 1980,4400 \right\} , \cdots ,\\
 \! & \left\{ 1620,5000 \right\}\left. \! \right\} \, \rads.
\label{eq:FreqGroup_Blindtests}
\end{split}
\end{equation}
The inversion of a frequency group is terminated after $20$ iterations or if the relative reduction of the misfit $\chi_{i}$ falls below a threshold value. As described in \autoref{subsec:inverseModeling}, a preconditioning of the gradient is performed.
In a distance of $2.5\,\m$ to the source and receiver stations as well as in a distance of $1.75\,\m$ to the free surfaces the gradient is set to zero. A transition zone of $2.5\,\m$ for the source and receiver stations and of $1.75\,\m$ for the free surfaces is applied where the gradient is increased from zero to is actual value.

\begin{figure}
\centering
\includegraphics[width=\textwidth]{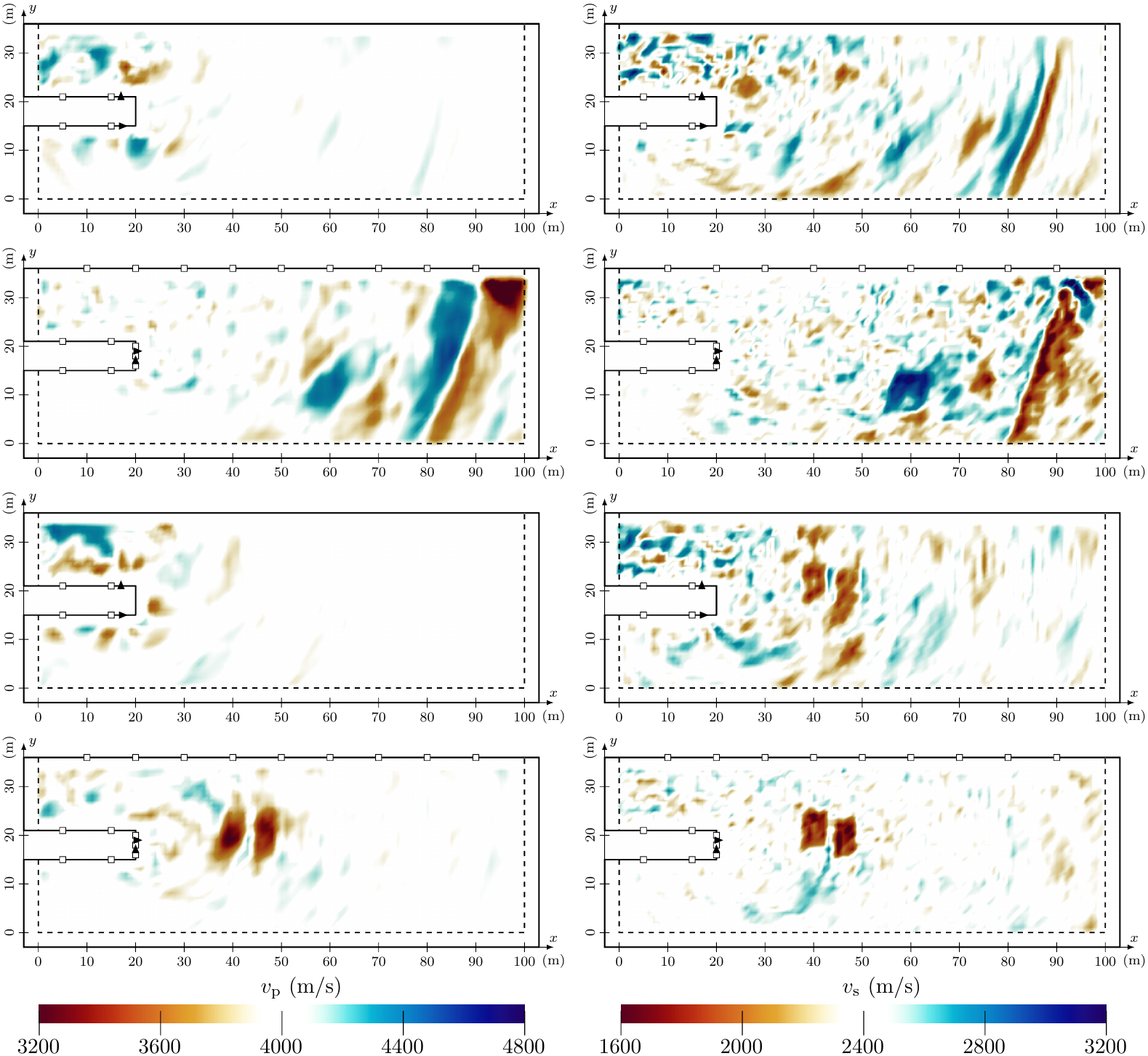}
\caption{P-wave (left) and S-wave velocity (right) distribution after the full waveform inversion. The first two rows correspond to the first investigated ground scenario, whereas the third and forth row belong to the second ground scenario. The used source and receiver stations are indicated in the same manner as in \autoref{fig:SR_Setup}.}
\label{fig:BlindtestsResultsFWI}
\end{figure}

The inversion results for both ground scenarios as well as for both source and receiver station configurations are illustrated in \autoref{fig:BlindtestsResultsFWI} by a visualization of the P-wave velocity as well as the S-wave velocity distribution over the used tunnel domain.

The resulting P-wave velocity field for the first ground scenario and the first source and receiver station configuration (\autoref{fig:BlindtestsResultsFWI} top row left) does not allow any prediction because most changes occur around the tunnel and seem to be spurious fluctuations. The S-wave velocity field is more sensitive for geological changes because the resolution of the wave velocities are proportional to their wave lengths. Therefore, the resulting S-wave velocity distribution (\autoref{fig:BlindtestsResultsFWI} top row right) allows a better prediction. A rapid increase followed by and sudden decrease can be determined approximately along a line from $\left( x=78.5,y=0 \right)$ to $\left( x=91,y=32 \right)$. This phenomenon has already been observed by \citet{riedel2021} where it belonged to a layer change. Due to the ambiguity of the inverse problem, the full waveform inversion tries to reconstruct geological features which produce the same reflections. The increase in front of the decrease leads to the same gradient of the ground properties and the same reflection behavior at an interface like a sudden change of the material properties. Therefore, it is probable that this feature belongs to a layer change where the S-wave velocity is decreased by approximately $891\,\ms$. If the interface continuous up to the surface can not be predicted, but by tacking the used preconditioning of the gradient into account it seems likely. Many fluctuations occur between the Earth's surface and the tunnel ceiling, but they can be interpreted as spurious fluctuation, due to the inaccuracies of the frequency domain wave modeling in this region. Some additional changes of the S-wave velocity can be observed, but a distinction between real features or spurious fluctuations is not possible because they are too unincisive.

The results of an application of the second source and receiver station configuration for the inversion of the seismic records of the first ground scenario allow an improved prediction (\autoref{fig:BlindtestsResultsFWI} second row). The indications for a layer change can be clearly identified in both wave velocity fields. In means of the S-wave velocity field the position of the layer change would be predicted along a line from $\left( x=77.75,y=0 \right)$ to $\left( x=91.75,y=36 \right)$ with an approximately decrease of the P-wave velocity of $946\,\ms$ and of the S-wave velocity of $1117\,\ms$ (evaluated at approximately the middle of the line).
A second disturbance with a circular shape at (x=59.5,y=10.5) with a radius of $4\,\m$ is detectable in both property fields. The object has an increased stiffness with the maximum values $\vp = 4427\,\ms$ and $\vs = 3105\,\ms$. Many other small increase and decreases of the wave velocities occur around the two predicted disturbances but are treated as spurious fluctuations because they do not appear in both velocity fields at the same position. Nevertheless, these fluctuations may impair the accuracy of the other predictions.

The application of the first source and receiver station configuration for the inversion of the seismic records of the second ground scenario leads again to a P-wave velocity distribution (\autoref{fig:BlindtestsResultsFWI} third row left) which gives no hints about the ground conditions in front of the tunnel.
The S-wave velocity field (\autoref{fig:BlindtestsResultsFWI} third row right) allows a rough guess of the geology in front of the tunnel.
A small rectangular disturbance (approximately $4\,\m \times 6\,\m$) is identifiable $17.5\,\m$ in front of the front tunnel face in front of a second bigger rectangular disturbance (approximately $4\,\m \times 10\,\m$) which have both a decreased S-wave velocity of approximately $\vs \approx 2000\,\ms$. The first disturbance is located on the level of the tunnel's ceiling and the second disturbance is located approximately on the level of the tunnel's center line. Below the second disturbance is another conspicuous decrease of the S-wave velocity to approximately $\vs \approx 2000\,\ms$ which is smaller in size. It might be another disturbance but could only be a spurious fluctuation, too. 
Comparatively many spurious fluctuation occur again between the Earth's surface and the tunnel's ceiling.

The utilization of the second source and receiver station configuration leads again to an improved image of the ground properties (\autoref{fig:BlindtestsResultsFWI} fourth row). The two disturbances are precisely identifiable, especially in means of the S-wave velocity distribution.
The first disturbance is located $17.5\,\m$ in front of the tunnel and $11\,\m$ below the Earth's surface with $\vp \approx 3535\,\ms$ and $\vs \approx 1899\,\ms$ whereas the second disturbance is located $2\,\m$ behind the first disturbance and $13\,\m$ below the Earth's surface with $\vp \approx 3454\,\ms$ and $\vs \approx 1802\,\ms$.
Both have approximately an area of $4\,\m\times7\,\m$.
The third velocity decrease, which is imaged by the resulting S-wave velocity field of the inversion with the other source and receiver station configuration, is no longer visible and therefore, it is assumed to be a spurious fluctuation.

\begin{figure}
\centering
\includegraphics[width=\textwidth]{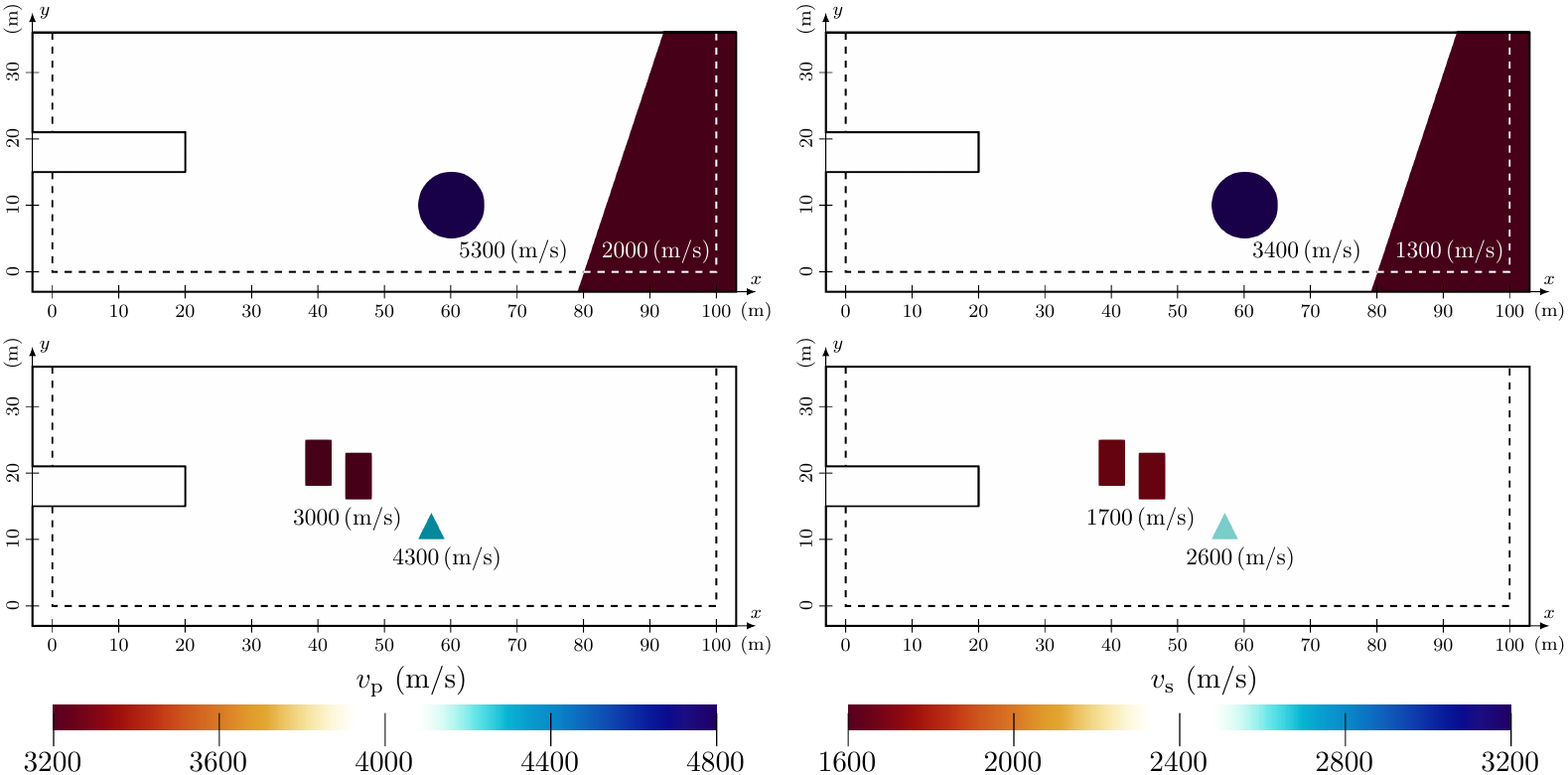}
\caption{P-wave (left) and S-wave velocity (right) distribution which have been used to produce the seismic records of the presented blind tests. The first ground scenario is displayed on the top and the second ground scenario is displayed at the bottom.}
\label{fig:BlindtestsReal}
\end{figure}

The ground properties which have been used to produce the seismic records of the blind tests have been revealed after the application of the full waveform inversion and are illustrated in \autoref{fig:BlindtestsReal}.
The prediction of the position of the layer change of the first ground scenario has been quiet accurate for both source and receiver station configuration, whereby some uncertainty has been left for the first source and receiver station configuration, because the P-wave velocity field gave no hints about the ground properties. The decrease of the S-wave velocities for the first configuration was not as high as in the real model but sufficient for prediction purposes. The wave velocity decreases for the second configuration were more accurate. The P-wave velocity decrease was just around half of the real one, but the S-wave velocity decrease was nearly the same as the real one.  
The circular disturbance appears to be slightly indicated by the S-wave velocity distribution of the first configuration by having the reference ground conditions in mind.
The position of the circle is predicted very accurately and the radius is predicted a little bit smaller by the second configuration. The predicted increase of the P-wave velocity is just a third of the increase of the reference model, but the increase of the S-wave velocity is predicted comparatively close the increase of the reference ground model.  

Within the reference ground model of the second ground scenario two rectangular disturbances with $\vp = 3000\,\ms$ and $\vs=1700\,\ms$ are located in front of the tunnel and have a slightly vertical offset to each other. A third disturbance with a small rectangular shape is located right underneath the second disturbance. With $\vp=4300\,\ms$ and $\vs=2600\,\ms$ the wave velocities are very slightly increased in comparison to the ambient ground properties. 
A prediction of the first and second disturbance has been able by both configurations of source and receiver stations. The prediction of the size, shape and values is enhanced for the second configuration. Especially, the predicted S-wave velocity is very close to the one of the reference model and less spurious fluctuations occur. The triangular shaped disturbance is not captured by any inversion result. On the one hand, the size of the object and the deviation of the wave velocities from the ambient values is so small that even if it would be indicated by the inversion results it might be mixed up with some spurious fluctuations. On the other hand, the disturbance is hidden behind the other two disturbances and is located underneath the tunnel. Therefore, less waves which are reflected at its material interfaces are gathered by the receiver stations and nearly no waves which are refracted by this disturbance are captured by the receivers.

In summary, the changes of the elastic properties are usually underestimated by the P-wave and S-wave velocity fields which are reconstructed by the full waveform inversion.
Spatial extensive disturbances as well as disturbances right in front of the front tunnel face are usually detectable by using only source and receiver stations around the tunnel. Additional receiver stations at the Earth's surface improve the level of detail of the ground image, especially for the P-wave velocity. The prediction of the position, shape and properties of a disturbance is significantly improved if not only waves are captured by the receiver stations which have been reflected by the disturbance but also waves which have been refracted by the material discontinuities. Small disturbances with properties that just slightly differ from the ambient properties might not be detectable in the presence of more prominent ground changes.
\section{Elastic three-dimensional layer change}
\label{sec:layerChange3D}

Since only two-dimensional tunnel domains are investigated in the previous section, this section analyzes the application of full waveform inversion on synthetic seismic records from a three-dimensional tunnel domain, which contains a layer change, using a three-dimensional model for the inversion process, too.

Two-dimensional models enable a fast computation of the seismic wave fields. This gives the opportunity to identify inversion parameters and investigate different inversion strategies quickly. Additionally, a first assessment can be done whether the application of full waveform inversion in the context of mechanized tunneling is a promising approach or not.
When it comes to an application on seismic records of a three-dimensional domain some characteristics of the used two-dimensional models have to be discussed.

For two-dimensional models symmetry in the third spatial direction is assumed. Therefore, a two-dimensional tunnel model actually represent a domain with a rectangular tunnel and a line load both with an infinite width in the third spatial direction. 
This leads, on the one hand, to a nearly complete separation of the waves below and above the tunnel and, on the other hand, to the phenomenon that seismic waves are continuously reflected at the tunnel's ceiling and the Earth's surface, which has already been discussed in \autoref{subsec:validationTunnelDomain}. 
The curved surface of the cylindrical tunnel allows additional radial tunnel surface waves.
The slope of material discontinuities in the third spatial direction is neglected by the symmetry assumption. The reflection and refraction behavior of the sloped interfaces can not be completely reconstructed by a full waveform inversion approach that uses a two-dimensional model.

Difficulties are caused by using unmodified seismic records of a three-dimensional domain where a point source is applied for an inversion with a two-dimensional model. Since the two-dimensional model assumes a line load along the third spatial direction, the amplitudes as well as the phase shifts of the simulated waves are changed due to the cylindrical geometrical spreading instead of the spherical geometrical spreading of the waves.

In many applications different modifications of the seismic records are performed to circumvent that the differences of measured data and modeled waveforms inhibit an application of a two-dimensional model.
The Bleistein filter \citep{bleistein1986} and a simplified version of it -- the so called $\sqrt{t}$-filter (cf. \citet{bretaudeau2013}) -- are often used to deal with this issue. Since the Bleistein filter is derived from a comparison of the two-dimensional and the three-dimensional analytical solution of the acoustic wave equation in a homogeneous unbounded domain, the filter works still good for elastic body waves but faces difficulties for surface waves and ground inhomogeneities. An extensive analysis of the Bleistein filter and on its effects on full waveform inversion schemes was performed by \citet{auer2013}. In a (shallow) tunnel domain surface waves are dominant within the seismic records. Therefore, using only such a filter for transforming two-dimensional into three-dimensional seismic records seems not to be appropriate.  

\citet{bharadwaj2017} receive promising results by using the $\sqrt{t}$-filter for the amplitude correction in combination with an simultaneous estimation of source filters as well as receiver coupling factors for the phase correction during a time domain full waveform inversion with a two-dimensional model in a study which is associated with mechanized tunneling.

Nevertheless, an accurate and physically meaningful forward wave model, which is able to reproduce the waveforms from field observations, is necessary for seismic exploration in means of full waveform inversion. Therefore, an investigation whether an application of full waveform inversion with a three-dimensional model for exploration in mechanized tunneling would be advantageous seems crucial even if a real time application will at least not become feasible in the next years.

\subsection{Three-dimensional tunnel environment}
\label{subsec:tunnelDomain3D}

\begin{figure}
\includegraphics[width=\textwidth]{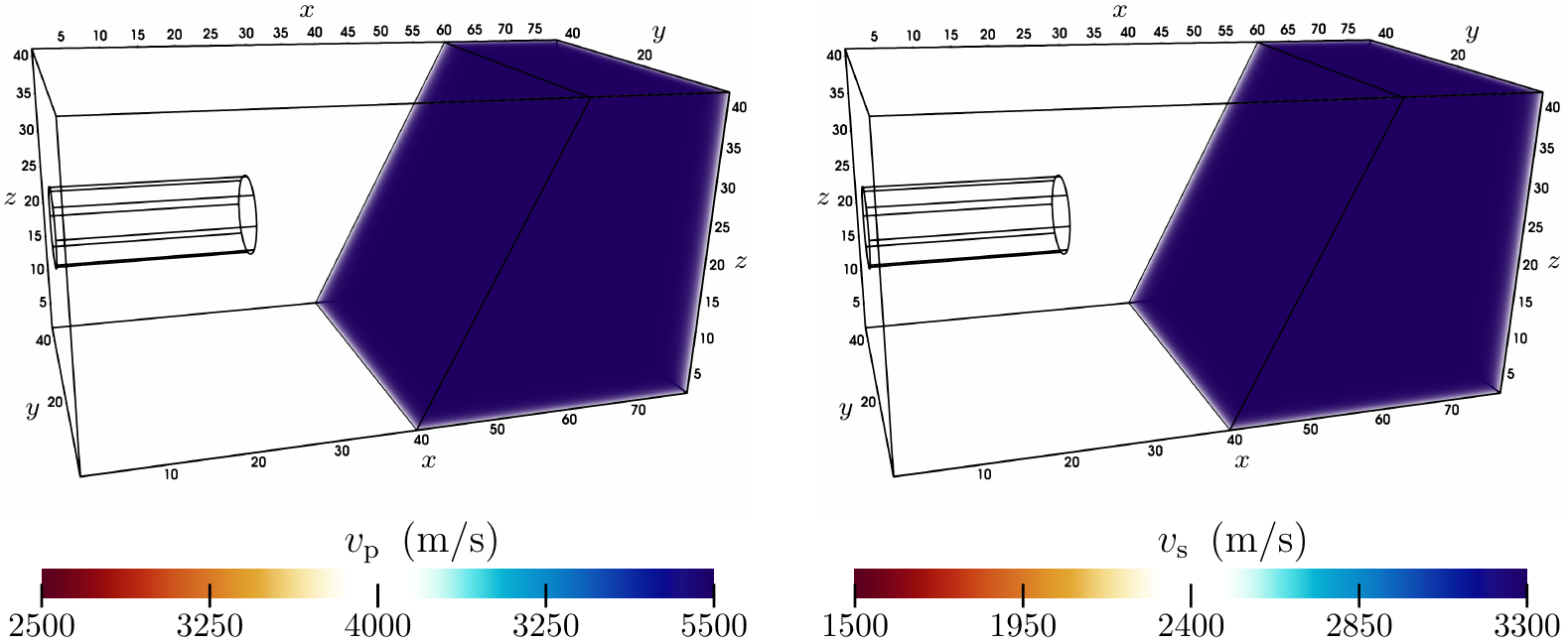}
\caption{P-wave (left) and S-wave velocity distribution (right) of the used synthetic reference ground model which contains an inclining layer change.}
\label{fig:Synthetic_SpecFEM_Fault}
\end{figure}

The dimensions of the used shallow tunnel domain as well as the layer change within the reference ground model are illustrated in \autoref{fig:Synthetic_SpecFEM_Fault}.The Earth's surface is located at the upper end of the $z$-axis. The tunnel has a diameter of $10\,\m$ and its center line is located $18\,\m$ below the Earth's surface. PMLs, which are not hidden in \autoref{fig:Synthetic_SpecFEM_Fault} but are not individually indicated, are applied at all other borders. Therefore, only the last $20\,\m$ of the already constructed tunnel is represented within the non absorbing domain. The ambient ground properties are $\vp = 4000 \, \ms$, $\vs = 2400 \, \ms$ and $\rho = 2500 \, \kgm$ which are used for the initial ground model, too.
The front of the second layer starts at $\left( x = 40 \, \m , z = 0 \, \m \right)$ and increases in an angle of $64^{\circ}$ to $\left( x = 60 \, \m , z = 41 \, \m \right)$. The plane of the interface is not rotated around the $z$-axis. The wave velocities of the second layer are increased compared to ambient values with $\vp = 5500 \, \ms$ and $\vs = 3300 \, \ms$, whereby the density is chosen constant.

For the frequency domain model a combination of prismatic and hexagonal elements with hierarchical higher-order shape functions \citep{zaglmayr2006} is used to take advantage of the preferable numerical properties of hexagonal elements in regular regions and, on the other hand, to have the possibility to discretize the tunnel track in an appropriate way without increasing the ratio between maximum and minimum element size by using prismatic elements.
The whole model employes $35\,062$ hexagonal elements as well as $9158$ prismatic elements where all PMLs have a width of $3$ elements.

Three different configurations of source and receiver stations are used for the inversion. The first configuration employes a source at the center of the tunnel front face which emits a signal in an $45^{\circ}$ angle into the ground. Therefore, a combination of compressive and shear waves are propagating perpendicular to the tunnel face. Four receiver stations are located at the front tunnel face as well as twelve receiver stations are evenly positioned around the tunnel walls.
The second configuration extends the fist configuration by using $18$ additional receiver stations at the Earth's surface. They are arranged along two lines in $x$-direction which have a distance of $8\,\m$ in y-direction to the tunnel center line. The lines start at $x=17.5\,\m$ and end at $x=57.5$. Therefore, the distance between each receiver station is $5\,\m$ and all receivers are in front of the second layer.
For the third configuration only a second source station is added to the second configuration at the Earth's surface over the center line of the tunnel at $x=40\,\m$ which emits a signal perpendicular to the Earth's surface into the ground.
All receivers are recording the displacements in all three spatial directions.
The position of the source and receiver stations as well as the excitation direction of the source stations are additionally indicated within the resulting P-wave and S-wave velocity distribution after the full waveform inversion in \autoref{fig:Results_SpecFEM_Fault}.

The seismic records of the reference tunnel domain are generated using SPECFEM3D Cartesian. For both source stations a Ricker wavelet with a peak frequency of $300\,\Hz$ without a time delay is used.

\subsection{Validation of three-dimensional elastic wave propagation in a tunnel domain}
\label{subsec:validationTunnelDomain3D}
The numerical PML parameter is evaluated by using the three-dimensional analytical frequency domain solution for a homogeneous unbounded domain which is given, inter alia, by \citet{gosselin2014}. Therefore, all displacements along a diagonal line are analyzed for an unbounded domain which has the same ground properties as well as comparable dimensions and uses a comparable discretization as the introduced tunnel model. With $c_{\mathrm{pml}} = 25\,000$ an appropriate damping behavior is adjusted for an angular frequency of $\omega = 1500\,\rads$.

\begin{figure}
\centering
\includegraphics[width=0.495\textwidth]{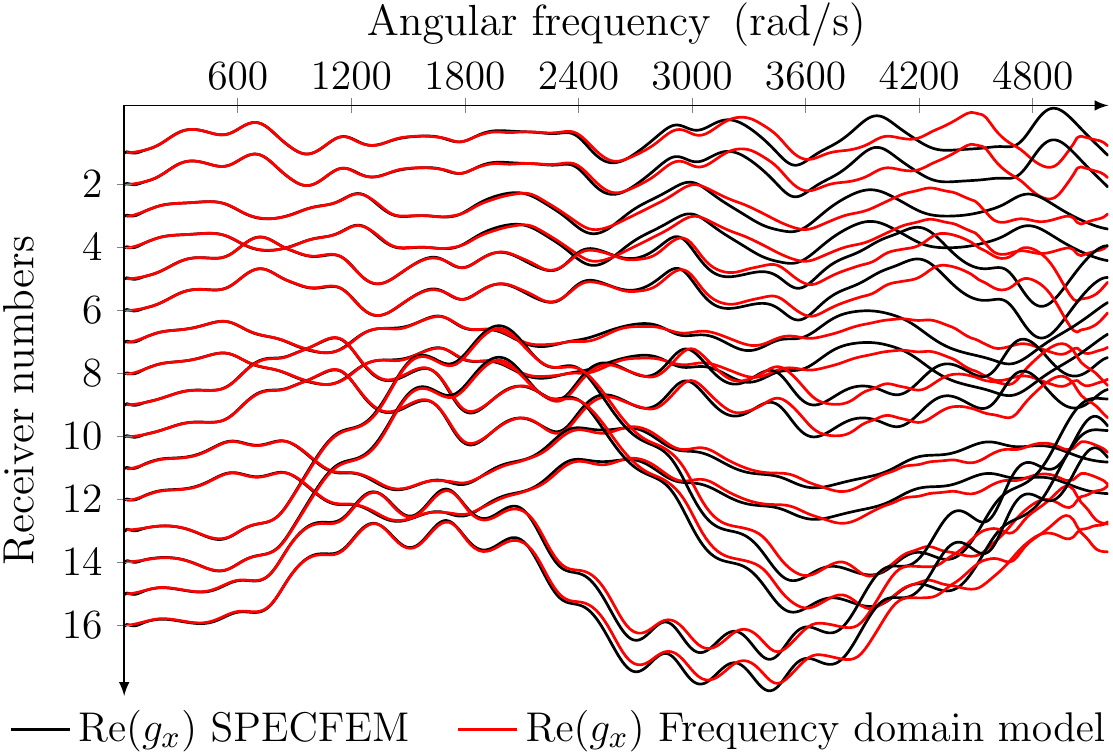}
\caption{Comparison of the real part of the frequency domain representation of the Green's functions in $x$-direction at the receiver station of the first source and receiver station configuration where the source and receivers are only located at the tunnel.}
\label{fig:ValidationTunnelDomain3D}
\end{figure}

The degree of the hierarchical shape functions for the wave modeling of the used tunnel domain is limited to $p=2$ due to limitations of the used computational resources. As a result only the wave fields of low frequencies are able to be computed with a sufficient accuracy.
An inverse discrete Fourier transformation of frequency domain seismic records with inaccurate values for higher frequencies would lead to erroneous time domain seismic records. For this reason, a validation of the wave modeling in a three dimensional tunnel domain is not possible by a comparison of time domain seismic records like in \autoref{subsec:validationTunnelDomain}.
Therefore, the frequency domain Green's functions are compared for the introduced tunnel domain but without any material discontinuities. The time domain seismic records, which are generated by SPECFEM3D Cartesian, are transformed into the frequency domain and a deconvolution with the used Ricker wavelet is performed to obtain the reference Green's functions.
The comparison of the real part of the frequency domain Green's functions in $x$-direction for the first source and receiver station configuration is illustrated in \autoref{fig:ValidationTunnelDomain3D}.
The receiver stations $1$--$4$ are located at the tunnel walls at $x=15\,\m$ in an angle of $45^{\circ}$, $135^{\circ}$, $225^{\circ}$ and $315^{\circ}$ to the $y$-$z$-axis.  
The receiver stations $5$--$8$ and $9$--$12$ are positioned in the same way but at $x=19\,\m$ and $x=23\,\m$. The receiver stations $13$--$16$ are located on the front tunnel face under the same angles with a distance of $3.5\,\m$ to the center of the tunnel face.
The frequency domain seismic records fit quite good up to a frequency of approximately $2000\,\rads$ but for higher frequencies the differences start to increase continuously. This allows only the use of low frequencies for the inversion procedure.

\subsection{Three-dimensional full waveform inversion}
\label{subsec:inversionResultsTunnelDomain3D}

While the used source function covers a frequency range between approximately $150\,\rads$ and $5000\,\rads$, only frequencies up to $1800\,\rads$ are used for the inversion procedure due to the given limitations of computational power.
Therefore, only a coarse resolution of the elastic properties after the inversion is assumed which might be sufficient for the investigation of big disturbances like the used layer change.
The first nine frequency groups only contain single frequencies, whereas the last four frequency groups combine a comparatively low frequency with a higher frequency:\\
\begin{equation}
\begin{split}
G = \left\{ \right. \! & \left\{ 200 \right\}, \left\{ 300 \right\}, \cdots , \left\{ 1000 \right\},\\
 \! & \left\{ 700,1200 \right\}, \left\{ 900,1400 \right\},\\
 \! &  \left\{ 1100,1600 \right\}, \left\{ 1300,1800 \right\} \left. \! \right\} \, \rads
\label{eq:FreqGroup_SpecFEM_Fault}
\end{split}
\end{equation}
A maximum number of ten iterations for every frequency group is used. The same preconditioning of the gradient is performed like for the blind tests (\autoref{subsec:resultsBlindtests}).

\begin{figure}
\includegraphics[width=\textwidth]{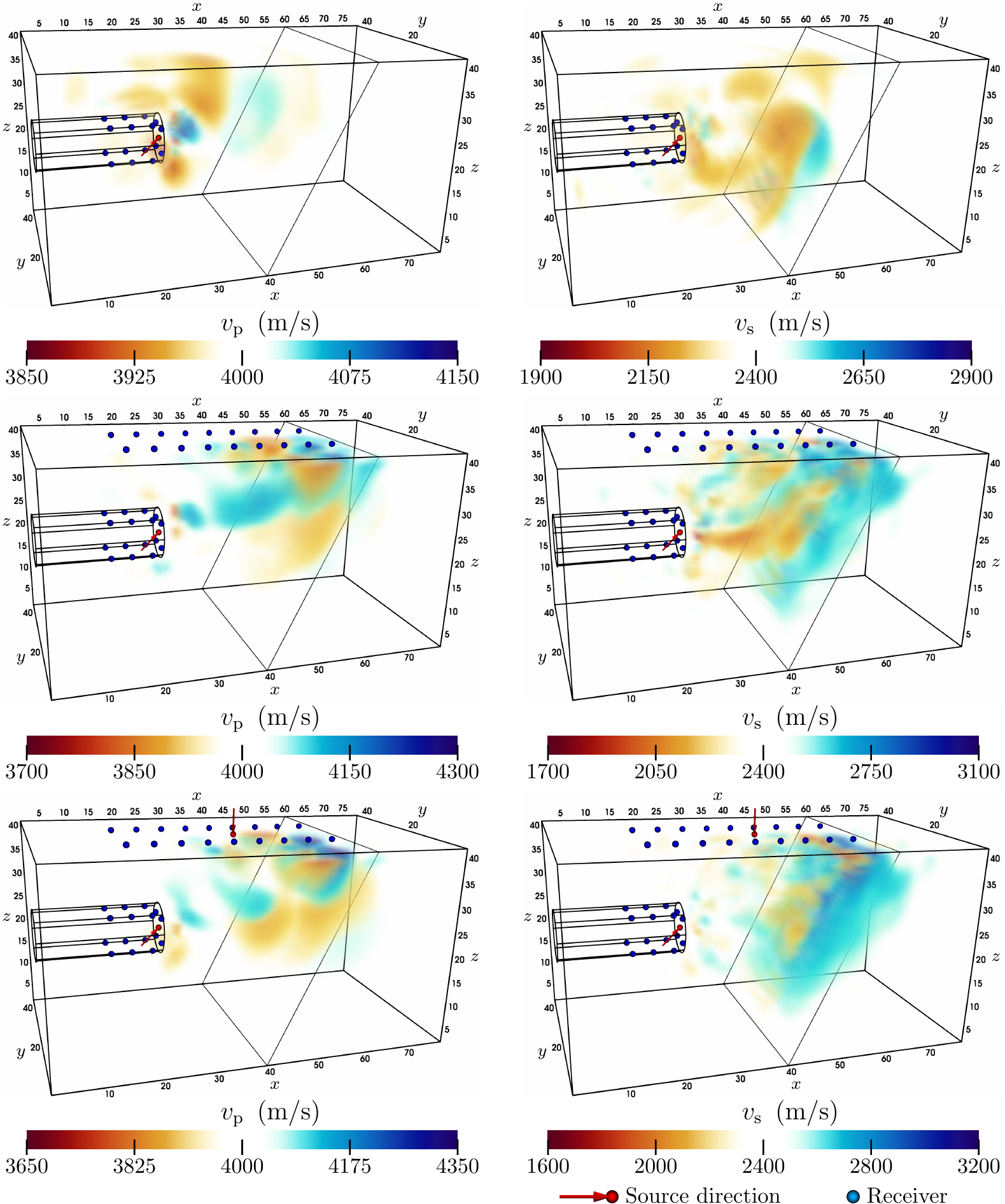}
\caption{The resulting P-wave (left) and S-wave velocity distribution (right) after the full waveform inversion. The locations of source and receiver stations are indicated for the three different configurations, which are illustrated line by line. The position of the searched layer change is indicated, too.}
\label{fig:Results_SpecFEM_Fault}
\end{figure}

The resulting P-wave and S-wave velocity distributions after the full waveform inversion are illustrated in \autoref{fig:Results_SpecFEM_Fault}.

The inversion results of the first configuration allow only a rough prediction of the geology in front of the tunnel face which was expected for the use of one source and receivers only around the tunnel. The P-wave velocity only changes rapidly in front of the tunnel (in front of the zone where the gradient was preconditioned). This phenomenon of erroneously high changes of the wave velocities around the source and receiver stations has been observed for the two-dimensional blind tests, too. Within the S-wave velocity field many erroneous fluctuations occur between the tunnel and the layer change, but at the interface between the two layers a decrease followed by a rapid increase on the level of the tunnel center line is observed. Such a behavior has already been observed for the layer change of the two-dimensional blind tests and therefore, a material change at this position can be predicted but neither is actual shape nor its dimensions.
The wavelike change of the S-wave velocity leads to a contrast of approximately $570\,\ms$. The increase of the S-wave velocity of the reference model at the interface of both layers is $900\,\ms$. Therefore, the predicted S-wave velocity increase amounts only two thirds of the actual one.

An interpretation of the P-wave velocity distribution of the second configuration (\autoref{fig:Results_SpecFEM_Fault} second row left) is quite difficult. On the one hand, an increase between the tunnel and the layer interface is observed and, on the other hand, many changes around the layer change occur. But these changes do not allow a prediction of the angle or the P-wave velocity of the second layer. A promising change is only observed directly underneath the Earth's surface. The reconstructed S-wave velocity field (\autoref{fig:Results_SpecFEM_Fault} second row right) predicts the position of the interface plane very precisely. Many fluctuations arise again between the tunnel and the second layer, especially on the level of the tunnel center line. The increase of the S-wave velocity is still not as high as in the reference model.

The inversion with the third configuration (\autoref{fig:Results_SpecFEM_Fault} third row) leads to the best results. Many changes of the P-wave velocity (left) occur around the layer change. The position and velocity change of the layer change is illustrated in a good way with a maximum contrast of $590\,\ms$ directly underneath the Earth's surface next to the last receiver stations. In combination with the S-wave velocity distribution (right) an accurate prediction of the layer change is possible. Less erroneous fluctuations occur for the S-wave velocity field and the increase of the velocity is accurately observed at the layer interface of the reference model. The velocity decrease in front of the velocity increase is not as pronounced as in the previous examples. Next to the Earth's surface a maximum velocity difference of approximately $705\,\ms$ is observed which gets quite close to the velocity increase of the reference model.

For all three inversion results the fluctuations between the ceiling of the tunnel and the Earth's surface, which are observed by the two-dimensional inversion results, are significantly reduced for an application of a three-dimensional model for the simulation of the wave propagation.
An accurate prediction becomes more difficult, because the number of property coefficients are significantly increased for three-dimensional inversion which increases the number of geological formations that describe the same features of the seismic records.
The inversion results get better for improved source and receiver station configurations. An additional application of receiver stations at the Earth's surface directly above the tunnel track might additionally improve the illumination of the ground directly in front of the tunnel face and might reduce erroneous fluctuations.
Only comparatively low frequencies are used for the performed inversion. Adding higher frequencies would increase the level of detail of the reconstructed wave velocity distributions and erroneous long-wave fluctuations would be reduced which simplifies the interpretation of the inversion results.

Overall, the proposed full waveform inversion approach leads to results which allow promising predictions about the geology in front of a tunnel. The quality of the prediction scales with the applied source and receiver station configuration.
\section{Conclusion}
\label{sec:conclusion}
A full waveform inversion scheme, which operates in the frequency domain, is applied on synthetic seismic records of different mechanized tunneling domains to predict disturbances in front of the tunnel boring machine. A suppression of spurious reflections from the artificial borders is achieved by using convolutional perfectly matched layers for the simulation of the elastic wave propagation. The L-BFGS method, the discrete adjoint gradient method, a quadratic approximation of the objective function with three points for the line search as well as a multi-scale approach are combined for the proposed scheme. 

No attenuation effects are considered in this study.
An identification of the numerical parameter of the used perfectly matched layer approach is performed by using analytical solutions of homogeneous unbounded domains in advance.
The simulation of elastic wave propagation within the chosen tunnel domain is validated by a comparison of the calculated seismic records (in the time or frequency domain) with their counterparts which are computed by spectral elements codes SPECFEM2D or SPECFEM3D Cartesian.
All reference seismic records of this study are produced by utilizing SPECFEM2D or SPECFEM3D Cartesian which produce different numerical errors than the wave propagation approach which is used for the inversion. On the one hand, this makes the inversion more challenging and, on the other hand, the inversion procedure gets more realistic since the seismic records from field observation are usually provided in the time domain, too.

First, two-dimensional blind tests are performed in a shallow tunnel domain where only the seismic records, the ambient ground properties as well as the used source functions are provided. Seismic records from two different ground scenarios for two different source and receiver station configuration are inverted.
A configuration which employes only source and receiver stations around the tunnel is able to predict conspicuous obstacles in front of the tunnel face but is not able to predict their shapes or properties accurately. 
Additional receiver stations at the Earth's surface improve significantly the imaging of the ground because refracted waves captured in addition to reflected waves. The positions, shapes as well as properties of the disturbances are predicted very accurately. But obstacles which just slightly differ from the ambient ground properties are not detectable if other more dominant disturbances are next to them.

The drawbacks of using a two-dimensional tunnel model for an inversion of a tunnel environment, which includes a cylindrical shaped tunnel, are discussed and the consequent need to study the application of full waveform inversion with a three-dimensional tunnel model is explained.
Therefore, a three-dimensional reference shallow tunnel model which includes an inclining layer change is used for a three-dimensional full waveform inversion where the inversion results of three different source and receiver station configurations are compared.
The reconstructed velocity field describes the layer change of the reference ground model in a better way as the source and receiver station configuration improves.

Overall, full waveform inversion turns out to be a promising tool for seismic exploration in mechanized tunneling but is at the moment not applicable for real-time application during the excavation process. An application in the future is expected to be possible since computational technologies have enormously evolved in the last decades. 

Nevertheless, it has to be considered that only synthetic reference seismic records are used without any added noise or any small variations of the used properties. Since such idealized reference data are not close to reality, a study is needed on how vulnerable full waveform inversion in the context of mechanized tunneling is to noise, to slightly inaccurate initial properties of the ground model as well as to errors due to geometrical simplifications of the ground domain.

The source function which is supposed to be emitted by the source stations is in most applications known, but the effective induced signal usually differs due to the coupling of the source and receiver stations with the ground. Therefore, approaches for an estimation of the effective source signature during or prior to full waveform inversion for reconnaissance in tunneling have to be studied.

Within the current studies on the application of full waveform inversion in mechanized tunneling, the inversion is performed and analyzed for an explicit advance of the tunnel. A continuous monitoring of the ground during the tunnel advancement by full waveform inversion might lead to better results. Because, on the one hand, a tunnel advancement leads to different spacings between source and receiver stations as well as disturbances and therefore, different information of the reflected and refracted waves might be captured. On the other hand, the predictions of former inversions might help to generate an improved initial ground model.
\section*{Acknowledgements}

\noindent The authors gratefully acknowledge the funding by the German Research Foundation (DFG) [grant number SFB837/3-2018] within the Collaborative Research Center SFB 837 ``Interaction modeling in mechanized tunneling'' within the subproject A2 ``Development of effective concepts for tunnel reconnaissance using acoustic methods''.\\
We thank the Computational Infrastructure for Geodynamics (\href{http://geodynamics.org}{http://geodynamics.org}) which is funded by the National Science Foundation under awards EAR-0949446 and EAR-1550901.

\clearpage

\bibliographystyle{apalike}

\end{document}